\documentclass[aps,superscriptaddress,amsmath,amssymb,floatfix,twocolumn,showpacs,amsfonts,longbibliography,notitlepage,superscriptaddress]{revtex4}
\usepackage{xr}
\usepackage{times}
\usepackage[varg]{txfonts}
\usepackage{textcomp}
\usepackage{graphicx}
\usepackage{subfigure}
\usepackage{color}
\usepackage[colorlinks=true,citecolor=blue,urlcolor=blue,linkcolor=blue,hyperindex]{hyperref}
\usepackage{braket}
\usepackage{float}
\usepackage{overpic}
\usepackage{bm}
\usepackage{array}
\usepackage{multirow}
\usepackage{booktabs}
\usepackage{appendix}
\usepackage{tabularx}

\makeatletter
\newcommand{\fmarki}{*}
\newcommand{\fmarkii}{\ensuremath{\dagger}}
\newcommand{\fmarkiii}{\ensuremath{\ddagger}}
\newcommand{\fmarkiv}{\ensuremath{\mathsection}}
\newcommand{\fmarkv}{\ensuremath{\mathparagraph}}
\newcommand{\fmarkvi}{\ensuremath{\|}}
\newcommand{\fmarkvii}{**}
\newcommand{\fmarkviii}{\ensuremath{\dagger\dagger}}
\newcommand{\fmarkix}{\ensuremath{\ddagger\ddagger}}

\def\@fnsymbol#1{{\ifcase#1\or \fmarki\or \fmarkii\or \fmarkiii\or \fmarkiv\or \fmarkv\or \fmarkvi\or \fmarkvii\or \fmarkviii\or \fmarkix \else\@ctrerr\fi}}
\makeatother

\renewcommand{\fmarki}{\ensuremath{\dagger}}
\renewcommand{\fmarkii}{*}
\renewcommand{\fmarkiii}{*}
\renewcommand{\fmarkiv}{*}
\renewcommand{\fmarkv}{x$_5$}
\renewcommand{\fmarkix}{z$_9$}

\begin{document}
	
	\title{Sliding multiferrocity in van der Waals layered CrI\(_2\) }

	\author{Hui-Shi Yu}
    \email{The authors contributed equally to this work.}
    \affiliation{Center for Neutron Science and Technology, Guangdong Provincial Key Laboratory of Magnetoelectric Physics and Devices, State Key Laboratory of Optoelectronic Materials and Technologies, School of Physics, Sun Yat-Sen University, Guangzhou, 510275, China}

    \author{Xiao-Sheng Ni}
\email{The authors contributed equally to this work.}
	\affiliation{Center for Neutron Science and Technology, Guangdong Provincial Key Laboratory of Magnetoelectric Physics and Devices, State Key Laboratory of Optoelectronic Materials and Technologies, School of Physics, Sun Yat-Sen University, Guangzhou, 510275, China}
    \affiliation{Peng Cheng Laboratory, Frontier Research Center, Shenzhen, China}

	\author{Kun Cao}
	\email{caok7@mail.sysu.edu.cn}
	\affiliation{Center for Neutron Science and Technology, Guangdong Provincial Key Laboratory of Magnetoelectric Physics and Devices, State Key Laboratory of Optoelectronic Materials and Technologies, School of Physics, Sun Yat-Sen University, Guangzhou, 510275, China}
    
\date{\today}

\maketitle

\newpage

{\bf Understanding magnetoelectric coupling in emerging van der Waals multiferroics is crucial for developing atomically thin spintronic devices. 
Here, we present a comprehensive first-principles investigation of magnetoelectric coupling in orthorhombic CrI$_2$. Monte Carlo simulations based on DFT-calculated magnetic exchange interactions suggest a proper-screw helimagnetic ground state with a  N\'{e}el temperature consistent with experimental observations. A ferroelectric switching pathway driven by interlayer sliding is predicted, featuring a low switching energy barrier and out-of-plane ferroelectric polarization.
To quantitatively characterize the magnetoelectric effect in orthorhombic CrI$_2$ and its microscopic origin, we evaluate the spin-driven polarization using the paramagnetic phase as a reference alongside the magnetoelectric tensor method. The extracted spin-driven polarization aligns along the $z$-axis, with its origin dominated by the exchange-striction mechanism. Although in-plane components of the total polarization in the bulk vanish due to global symmetry constraints, each CrI\(_2\) single layer exhibits local electric polarization along the $x$ direction, arising from the generalized spin-current mechanism, which couples spin chirality to the electric polarization. As a result, we further predict that a proper-screw helimagnetic state may persist in monolayer CrI\(_2\), with its charity reversable by switching the in-plane electric polarization through applying external electric field, providing another promising candidate for electrical control of two-dimensional multiferroics.
}
	
\maketitle	
~\\
\noindent{\bf\large Introduction}\\
	
Multiferroic materials, are usually characterized by the coexistence and coupling of spontaneous electric polarization and long-range magnetic order~\cite{cheong2007multiferroics,ramesh2007multiferroics,picozzi2009first}, offering promising avenues for applications in spintronics and magnetic data storage~\cite{wang2009multiferroicity,bibes2008towards}. 
The rapid rise of layered van der Waals (vdW) multiferroics has further expanded opportunities for the design of high-performance electronic devices. Of particular interest is the recent finding that type-II multiferroic order can persist down to a single atomic layer in the vdW material NiI\(_{2}\)~\cite{song2022evidence}.
However, despite such progress, the scarcity of atomically thin multiferroics has driven theoretical efforts to elucidate spin-driven (SD) electric polarization~\cite{katsura2005spin,sergienko2006role,sergienko2006ferroelectricity,choi2008ferroelectricity,murakawa2010ferroelectricity,tokura2014multiferroics,xiang2013unified}, aiming to achieve strong magnetoelectric (ME) coupling in low dimensional materials.

\begin{figure}[t]
		\includegraphics[scale=0.25]{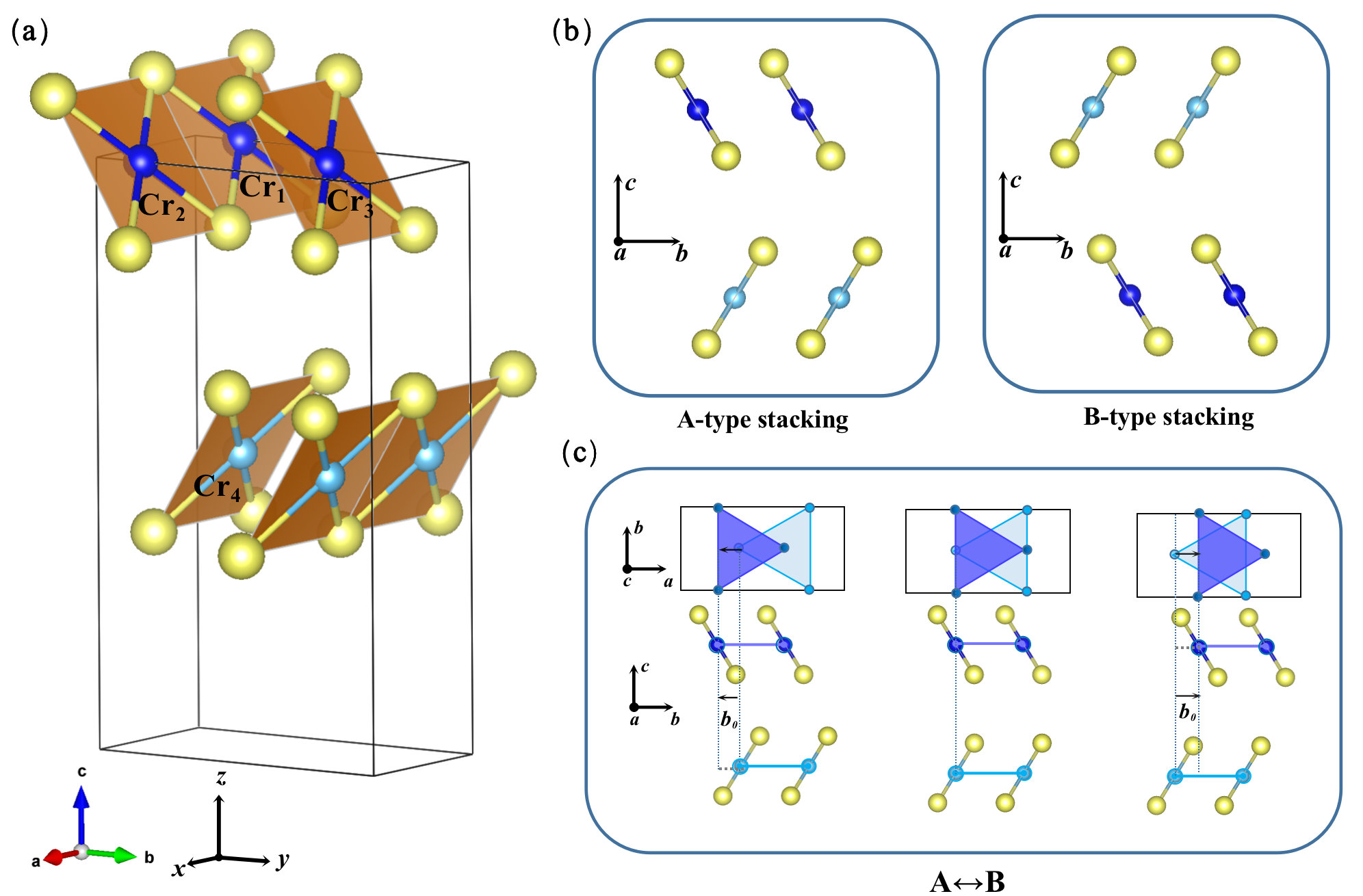}
		\caption{\textbf{Crystal structure of \textit{O}-phase CrI$_2$}.(a) Side view of the \textit{O}-phase CrI$_2$ crystal structure in a unit cell.
(b) A-type and B-type stacking configurations.
(c) Sliding transformations connecting the A-type and B-type stacking. Black rectangles outline unit cells.}
	\label{crystal}
	\end{figure}

The recently discovered van der Waals layered CrI$_2$ is another promising candidate for realizing atomically thin multiferroicity. 
However, limited bulk experimental studies on CrI\(_2\) have led to ongoing debate regarding its crystallization in either the monoclinic phase (\textit{M}-phase, space group \(C2/m\)) or the orthorhombic phase (\textit{O}-phase, space group \(Cmc2_1\))  ~\cite{zhang2022structural,besrest1973structure}.
Very recently, experimental measurements have verified the crystal structure of the orthorhombic phase~\cite{schneeloch2024helimagnetism}.
Regarding the magnetic ground state of this compound, antiferromagnetism ~\cite{zhang2022structural}, ferromagnetism ~\cite{peng2020mott}, and helimagnetism~\cite{schneeloch2024helimagnetism} have all been reported perviously.
Moreover, the \(O\)-phase is formed by a 180° rotation and a translational sliding between two single-layers[see Fig.\ref{crystal}], and such a unique interlayer stacking is also a signature of potential ferroelectricity.
Several theoretical studies suggest that the \(O\)-phase exhibits an out-of-plane polarization that can be reversed by sliding the layers ~\cite{zhang2022structural,acharya2024first}.
Sliding-induced polarization switching has also recently emerged as a key focus in sliding ferroelectricity, with great potential for high-density nonvolatile memory applications~\cite{wu2021sliding,bian2024developing,li2017binary}.
To the best of our knowledge, while the intrinsic ferroelectric polarization can be primarily attributed to the large off-center atomic displacements, the ME coupling, which can be characterized by the variation of the ferroelectric polarization induced by the onset of magnetic ordering and its underlying microscopic mechanism, remain elusive to date.
 
In this study, we investigate the magnetic structure and ME coupling in the \(O\)-phase of CrI$_{2}$. Monte Carlo (MC) simulations, combined with DFT-calculated magnetic exchange parameters, confirm a helimagnetic (HM) ground state with a  N\'{e}el temperature \(T_{\mathrm{N}}\) in agreement with experiments. Furthermore, we predict ferroelectric switching via interlayer sliding, characterized by a low-enengy barrier.
To elucidate the ME coupling in this type-I multiferroic, we analyze the spin-driven polarization \(P_{\mathrm{SD}}\), defined as the change in electric polarization induced by magnetic ordering. By simulating the paramagnetic (PM) phase and subtracting its polarization from that of the HM phase, we isolate a net \(P_{\mathrm{SD}}\) along the \(z\)-axis, quantitatively consistent with results from  our ME tensor analysis.
Both methods reveal that the spin-driven polarization primarily originates from the exchange striction mechanism. 
Interestingly, although the net \(x\)-component of the total polarization vanishes due to global symmetry constraints, individual layers produce local polarization along the \(x\)-axis via the generalized spin-current (GSC) mechanism, which can couple spin chirality with electric polarization.
Based on these insights, we predict that a proper-screw HM state may still persist in monolayer CrI$_{2}$ and its magnetic chirality can be switched by external electric field, facilitated by reversing the in-plane polarization.

~\\
\noindent {\bf\large Results and discussions}\\
\noindent {\bf\large Crystal structure}\\
	
Bulk CrI\(_2\) crystallizes in an orthorhombic phase (\textit{O}-phase, space group \textit{$Cmc2_1$} [$No.$ 36]), with experimental lattice parameters of \(a = 3.907\) \AA, \(b = 7.496\) \AA, and \(c = 13.479\) \AA~\cite{PhysRevB.109.144403}. The unit cell exhibits a layered architecture containing two CrI$_2$ monolayers, each consisting of a chromium layer sandwiched between two iodine sublayers.  Below the experimental N\'{e}el temperature \(T_N = 17\ \text{K}\), this phase hosts an incommensurate HM order with \(\boldsymbol{Q} = (0.2492,\ 0,\ 0)\)~\cite{schneeloch2024helimagnetism}. Notably, CrI\(_2\) hosts a competing monoclinic polymorph (\textit{M}-phase, space group \textit{$C2/m$} [$No.$ 12]). Neutron powder diffraction measurements reveal the coexistence of a dominant \textit{O}-phase with a minor component of \textit{M}-phase (about 1 vol\% at 30 K). Furthermore, no structural phase transition has been observed across the measured temperature range of 100–400K~\cite{schneeloch2024helimagnetism}. Based on these findings, we focus our investigation on the \textit{O}-phase CrI\(_{2}\). 

As shown in Fig.~\ref{crystal}(a), a CrI\(_{2}\) ribbon is oriented along the \(a\)-axis, with its planes parallel to the \(ab\) plane.
In bulk CrI\(_2\), Cr-Cr bonds can be categorized into three types based on their spatial relationships with the layers and ribbon chains: intrachain, intralayer interchain, and interlayer.  
The nearest-neighbor Cr-Cr bonds, Cr$_{1}$-Cr$_{2}$ (intrachain), Cr$_{1}$-Cr$_{3}$ (intralayer interchain), and Cr$_{1}$-Cr$_{4}$ (interlayer), are illustrated in Fig.~\ref{crystal}(a).
For the \(O\)-phase CrI\(_2\), the CrI\(_6\) octahedra in the upper and lower layers are not only tilted in different directions, but also exhibit a relative translational sliding [see Fig.\ref{crystal}(a)].
This stacking configuration is not centrosymmetric and breaks mirror symmetry along the \textit{z}-axis. Therefore, it is potentially ferroelectric with spontaneous polarization parallel to the \textit{z}-axis, although no experimental confirmation is reported so far.

Two distinct stacking configurations of \(O\)-CrI\(_2\), A-type and B-type, are presented in Fig.~\ref{crystal}(b).
The transformation from the A-type to the B-type stacking can be characterized by alternating relative translation of adjacent layers along the \( b \)-axis by approximately \( \pm 0.34  b\) [see Fig.\ref{crystal}(c)].
Similar sliding transformations can be found in other layered vdW materials due to weak interlayer interactions\cite{constantinescu2013stacking,sivadas2018stacking}. In the case without spin polarization, the atomic structure of the B-type stacking are connected by inversion operation to the A-type stacking. Therefore, unless otherwise noted, we only discuss results calculated with the A-type stacking.

~\\
\noindent {\bf\large Magnetic properties}\\

\begin{figure}[t]
\centering
	\includegraphics[scale=0.45]{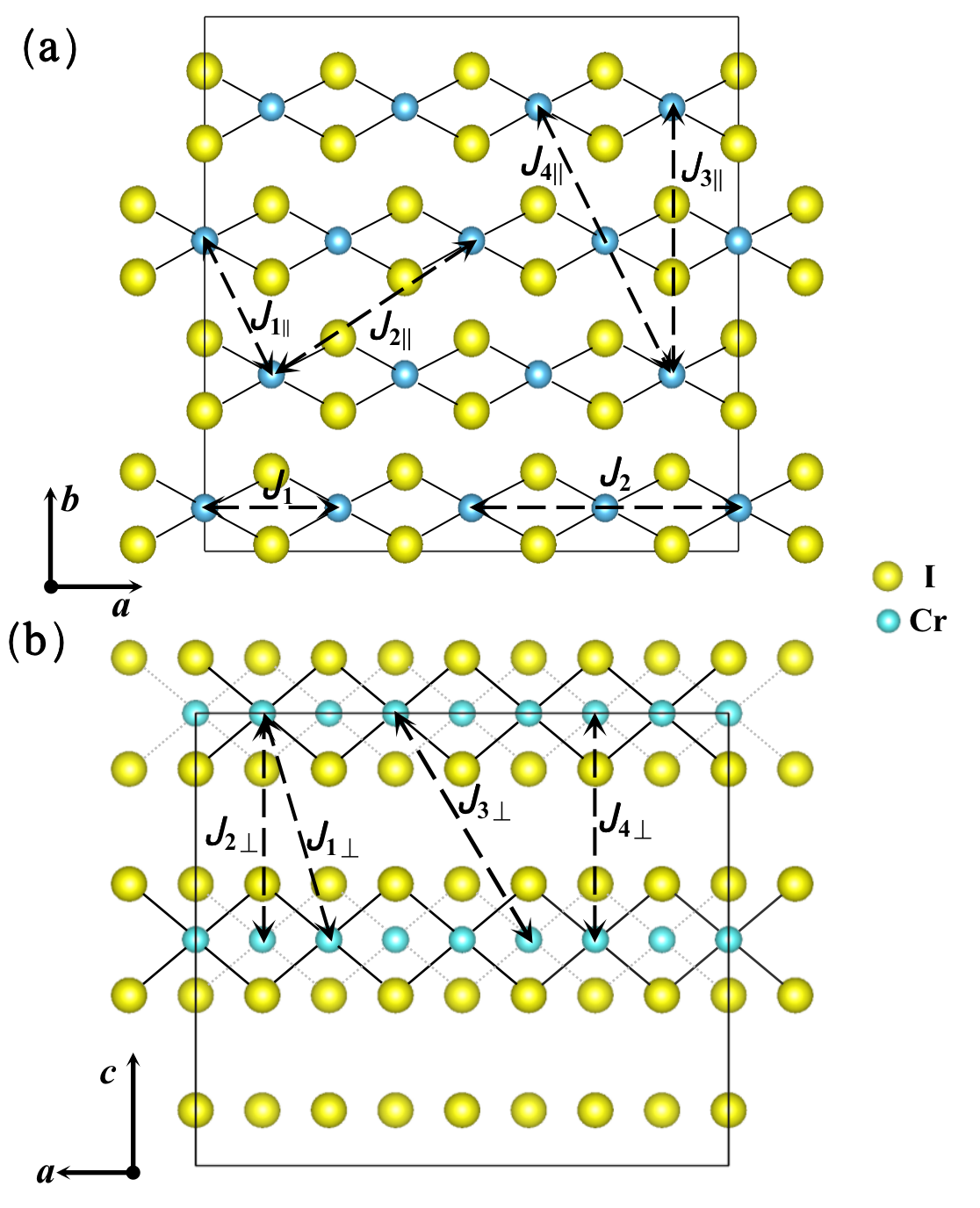}
	\caption{\textbf{Exchange paths in $O$-phase CrI$_2$}. Illustration of the exchange pathways in the $O$-phase CrI$_2$. The corresponding bond lengths and values of the interactions are listed in Table~\ref{tab2}.}
\label{Jpath}
\end{figure}

In order to model the magnetic properties of  CrI$_2$, we employ the classic Heisenberg model:
	\begin{equation}
        \textit{H} = \sum_{\langle i,j \rangle_n} J_{n} \mathbf{S}_i \cdot \mathbf{S}_j+\sum_{\langle i,j \rangle_n^\parallel} J_{n}^{{\parallel}} \mathbf{S}_i \cdot \mathbf{S}_j+\sum_{\langle i,j \rangle_n^\perp} J_{n}^{{\perp}} \mathbf{S}_i \cdot \mathbf{S}_j,
    \label{eq:1}
\end{equation}
where $\boldsymbol{S}_i$ and $\boldsymbol{S}_j$ are the spins at sites $i$ and $j$, respectively. $J_n$ represents the $n$th nearest neighbor (NN) Cr-Cr intrachain interactions. The superscripts $\parallel$ and $\perp$ indicate intralayer interchain interactions and interlayer interactions, respectively.
The calculated \( J \) values are summarized in Table~\ref{tab2}, with their exchange paths illustrated in Fig.~\ref{Jpath}.

We begin by considering a minimal model that includes only the three strongest exchange interactions: the intrachain \( J_1 \) and \( J_2 \), as well as the interlayer interchain \( J_1^{\parallel} \). The spin configuration can be expressed as:
\(
\mathbf{S} = S_0 \cdot \big[\cos\theta_1 \cos(\mathbf{Q} \cdot \mathbf{R}), -\sin(\mathbf{Q} \cdot \mathbf{R}), \sin\theta_1 \cos(\mathbf{Q} \cdot \mathbf{R})\big],
\)
where \(\theta_1\) denotes the angle between the \(z\)-axis and the normal to the rotation plane.
For a propagation vector \(\bm{Q} = (q_a, 0, 0)\) along the ribbon chains, the magnetic Hamiltonian incorporating only \( J_1 \), \( J_2 \), and \( J_1^{\parallel} \) is given by:

\begin{equation}
\begin{aligned}
H_{iso} &= J_{1} S_{0}^2\cos(q_a)+J_{2} S_{0}^2 \sin(2q_a) \\
&\quad +J_{1}^{\parallel}S_{0}^2(\cos(q_a)+1).
\end{aligned}
\label{eq:Hiso}
\end{equation}

By differentiating Eq. (\ref{eq:Hiso}) and setting \(\frac{\partial \textit{H}_{iso}}{\partial q_a} = 0\), we can determine the corresponding magnetic ground state via the relation:
	\begin{equation}
q_a \propto \arctan \left( \frac{J_1 + J_1^\parallel}{\sqrt{16 J_2^2 - (J_1 + J_1^\parallel)^2}} \right).
 \label{eq:qa}
\end{equation}
From Eq. (\ref{eq:qa}), we can deduce that HM arises if \( \frac{|J_2 |}{|J_1+J_1^{\parallel}|} > \frac{1}{4}\). Our calculated \(J_1\), \(J_2\), and \(J_1^{\parallel}\) satisfy these conditions (See Table~\ref{tab2}), thus suggesting that the ground state of CrI$_2$ is HM due to the magnetic frustration between the \(J_{1}\) and \(J_{2}\).

MC simulations are further performed to validate our predictions. Using the calculated exchange interactions (see Table~\ref{tab2}), our MC simulations confirm that the ground state is a noncollinear HM phase with a N\'{e}el temperature \(T_N\) of 16 K, in good agreement with the experimental \(T_N\) of 17 K~\cite{schneeloch2024helimagnetism}.  Fourier analysis of spin configurations obtained from MC simulations reveals that the magnetic ground state exhibits a wave vector \( q_a \) along the ribbon chain direction with \(|\boldsymbol{Q}| = (0.25,0,0)\), which is also consistent with experimental observations~\cite{schneeloch2024helimagnetism}.

\begin{table}[t!]
    \centering
    \caption{Exchange interactions and corresponding bond lengths for the \textit{O}-phase CrI$_2$ from DFT + $U$ calculations. We only consider exchange interactions with bond lengths less than 9 \AA\ .}
    \renewcommand{\arraystretch}{1.5} 
    \setlength{\tabcolsep}{10pt}
    	\begin{ruledtabular}
    \begin{tabular}{ccc}
    \toprule
       
      Type & Cr-Cr (\AA)  & $J_{n}$ (meV)   \\
        \hline
        \multirow{2}{*}{intrachain} & 3.91& $J_{1}$ = 1.25  \\
        &7.81 & $J_{2}$ = 1.15  \\
        \hline
        \multirow{4}{*}{intralayer interchain} & 4.23& $J_{1}^{\parallel}$ = 2.07  \\
       & 6.96 &$J_{2}^{\parallel}$ = -0.01 \\
       & 7.50 & $J_{3}^{\parallel}$ = 0.13\\
       & 8.45 &$J_{4}^{\parallel}$ = -0.58\\
        \hline  
         \multirow{4}{*}{interlayer}& 7.14  & $J_{1}^{\perp}$ = -0.34 \\
         & 7.16 &$J_{2}^{\perp}$ = 0.05 \\
         & 8.16 &$J_{3}^{\perp}$ = -0.14\\
         & 8.43 &$J_{4}^{\perp}$ = -0.15\\
    \toprule
    \end{tabular}
	\end{ruledtabular}
    \label{tab2}
\end{table}

\begin{figure*}[t]
\centering
	\includegraphics[scale=0.5]{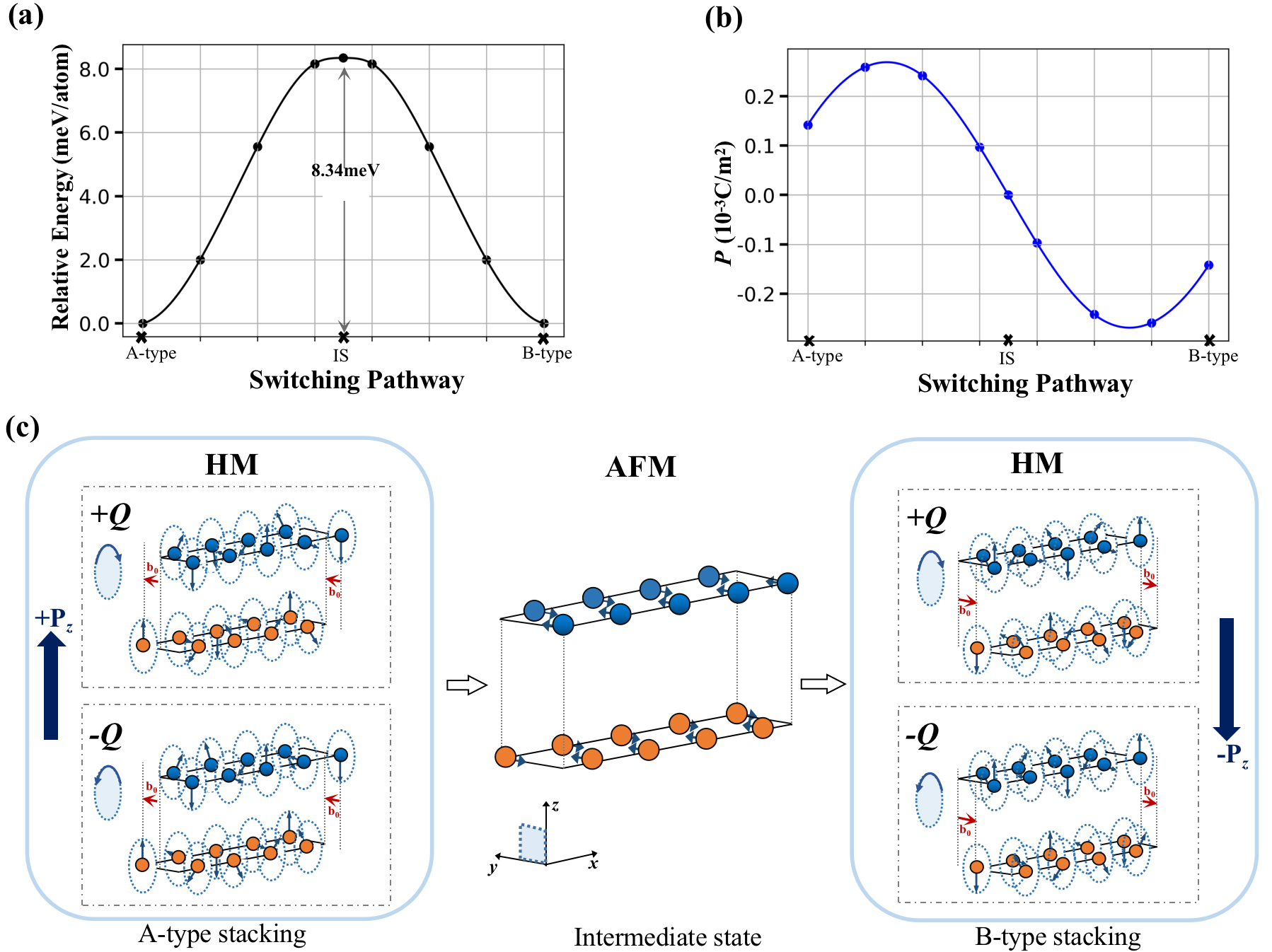}
	\caption{\textbf{Ferroelectric switching.} (a) (b) Energies and ferroelectric polarizations along the calculated switching path. (c) Illustration of the switching paths facilitated by interlayer sliding and its corresponding magnetic structures. Note that the spin spiral lies within the \(yz\) plane (shaded in blue).}
\label{stwitchingpath}
\end{figure*}

Moreover, we discover that the relative phase of the spin helices with respect to both adjacent in-plane chains and out-of-plane stacked chains is consistent with the description provided on the basis of experimental data.~\cite{schneeloch2024helimagnetism}.
Specifically, the interchain spin arrangement is determined to be AFM along the \( b \)-direction due to the AFM \( J_{1}^{\parallel} \). In contrast, the ferromagnetic (FM) \( J_{1}^{\perp} \) indicates FM coupling between adjacent interlayer chains.
To further distinguish between cycloidal and proper-screw HM orders, we perform DFT calculations with spin-orbit coupling (SOC) using a \(4 \times 1 \times 1\) supercell with \(\boldsymbol{Q} = (0.25, 0, 0)\) along the \(a\)-axis.
Our results indicate that the proper-screw HM state is energetically favored, with an energy 0.19 meV/Cr lower than that of the cycloidal state.  
As a result, we identify the magnetic ground state of the \(O\)-phase CrI\(_2\) as a proper-screw helimagnetic state, with the relative phases between adjacent chains clearly depicted in Fig.~\ref{stwitchingpath}(c)

~\\
\noindent {\bf\large Ferroelectric switching}\\

As confirmed by our DFT calculations, the A-type stacking with the \(\boldsymbol{Q}\) directed along \(+a\) and \(-a\) are energetically degenerate, so as for the B-type stacking.
Therefore, the \(O\)-phase CrI\(_2\) hosts four degenerate configurations, as illustrated in Fig.~\ref{stwitchingpath}(c).
If switchable between each other by an applied electric field, the A-type stacking and B-type stacking are expected to host out-of-plane spontaneous electric polarization $\mathbf{P}$, but with opposite directions. Therefore, we label the four distinct configurations as $\mathbf{P}$↑\(\boldsymbol{Q}\)↑, $\mathbf{P}$↑\(\boldsymbol{Q}\)↓, $\mathbf{P}$↓\(\boldsymbol{Q}\)↑, and $\mathbf{P}$↓\(\boldsymbol{Q}\)↓,  
where $\mathbf{P}$↑ denotes the polarization vector pointing along the \(+z\) direction, and \(\boldsymbol{Q}\)↑ indicates the spin chirality vector aligning along the \(+a\) direction.
Similar to the polarization switching mode in sliding ferroelectrics \cite{wu2021sliding,bian2024developing,li2017binary}, we expect 
\(P\) to be reversed upon interlayer sliding and vanish when relative displacement \(b_{o}\) reaches zero  [see the central inset of Fig.\ref{stwitchingpath}(c)]. The intermediate state (IS) corresponds to the centrosymmetric space group \textit{C}\textit{m}\textit{ce} (No. 64) and represents a paraelectric phase (PE)\cite{zhang2022structural}.
By computing the exchange interactions and performing MC simulations[see supplemental materials], we propose that the IS structure stabilizes in a collinear antiferromagnetic (AFM) phase, as shown in Fig.\ref{stwitchingpath}(c).

The ferroelectric switching pathway calculated using the CI-NEB method reveals an adiabatic path along which both the magnitude and direction of $\mathbf{P}$ can be continuously tracked. 
The calculated value of the polarization for the HM \(O\)-phase CrI\(_2\) is 0.142\ ~\(\mu\text{C}/\text{cm}^2\), in agreement with previous reports on bulk samples~\cite{zhang2022structural,shirodkar2014emergence,li2017binary}. As shown in Fig.~\ref{stwitchingpath}(a), the ferroelectric switching process involves overcoming an energy barrier of approximately \(\Delta = 8.34\)\,meV/atom (equivalent to 25.02\ meV per formula unit, f.u.),
suggesting that the switching is feasible under experimentally accessible electric fields.
For comparison, this switching barrier is one order of magnitude lower than that of ferroelectric PbTiO\(_3\) (200\,meV/f.u.)~\cite{ederer2006origin}, but higher than that of the representative sliding ferroelectric BN (4.5\,meV/f.u.)~\cite{li2017binary}.

~\\
\noindent {\bf\large Spin-driven polarization}\\

Following real physical process, in type-I multiferroics with co-existing ferroelectric orders already in the PM phase, the SD polarization can be characterized by the variation of total polarization opon magnetic ordering, i.e. $\mathbf{P}_{SD}$= $\mathbf{P}_{HM}$-$\mathbf{P}_{PM}$~\cite{johnson2014cabaco}. Based on our calculations, it is expected that structural distortion induced by interlayer sliding is the primary contributor to the total polarization. Therefore, quantifying the SD polarization in the $O$-phase CrI$_2$ pose a challenge, as it is masked by the presence of a dominating pre-existing ferroelectric polarization. 

Accordingly, we first calculate the electric polarization in the orthorhombic PM phase $\mathbf{P}_{PM}$ as a reference. In order to simulate the PM state, we employ the magnetic sampling method (MSM) within the framework of the Heisenberg model. The MSM method has been proven effective in modeling PM states in systems such as ZnCr\(_2\)O\(_4\)~\cite{PhysRevLett.96.205505}, CaBaCo\(_4\)O\(_7\)~\cite{johnson2014cabaco}, and BaFe\(_{2-\delta}\)S\(_{1.5}\)Se\(_{1.5}\)~\cite{PhysRevB.105.214303}. This method simulates the PM state by averaging multiple spin configurations to cancel out the energy contributed by relevant exchange interactions, thereby satisfying the condition \(\varphi = \sum_{k} \langle \mathbf{S}_{i}(k) \cdot \mathbf{S}_{j}(k) \rangle = 0\), where \(k\) indexes distinct spin configurations. In this sense, local ME polarizations are also canceled out in the PM phase, with the assumption that local polarizations are linear to $\mathbf{S}_{i}(k) \cdot \mathbf{S}_{j}(k)$. In our case, we consider in total ten inequivalent exchange interactions  $J$'s ($J_{1} \sim J_{2}$, $J_{1}^{\parallel} \sim J_{4}^{\parallel}$, and $J_{1}^{\perp} \sim J_{4}^{\perp}$) and construct four collinear spin configurations within a \(4 \times 2 \times 1\) supercell to simulate the PM phase (see Fig.(1) in the Supplementary Material, effectively canceling out all the accounted exchange interactions.

By averaging the polarization of the MSM spin configurations, the electric polarization of the PM phase are then obtained as \(\mathbf{P}_{\text{PM}}^{\text{PM}} = (0.048, -0.002, 0.174) \, \mu\text{C/cm}^2\), where the superscript PM and subscript PM denotes that the electric polarization are calculated in the simulated PM phase with atomic structure relaxed under the same magnetic phase. The polarization of the HM phase is calculated to be (0.047, -0.001, 0.133)\(\mu\text{C/cm}^2\). As a result, the total SD polarization \( \mathbf{P}_{SD} = \mathbf{P}_{\text{HM}}^{\text{HM}}-\mathbf{P}_{\text{PM}}^{\text{PM}}   = (0, 0.001, -0.041) \, \mu\text{C/cm}^2\).
Our results also indicate that the \(\mathbf{P}_{\text{SD}}\) is oriented opposite to the total polarization, similar to observations in the type I multiferroic BiFeO\(_3\)~\cite{lee2013negative}. We further analyze \(\mathbf{P}_{\text{SD}}\) by decomposing it into purely electronic \(\mathbf{P}_{\text{ele}}\) and ionic \(\mathbf{P}_{\text{ion}}\) components. By imposing the HM spin configuration (below \(T_{N}\)), onto the same atomic structure relaxed in the reference PM phase, we can isolate the pure electronic component \(\mathbf{P}_{\text{ele}}\)= (0, 0, -0.040) \(\mu\text{C/cm}^2\). 
The resulting \(\mathbf{P}_{\text{ion}}\) is found negligible, with an amplitude of 0.001 \(\mu\text{C/cm}^2\) along the same direction as \(\mathbf{P}_{\text{ele}}\). We therefore neglect the ionic contributions for simplicity in the following analysis.

An alternative solution for evaluating \(\boldsymbol{P}_{\mathrm{SD}}\) is to employ the ME coupling tensor, which can be calculated by the four-state mapping method\cite{wang2016microscopic}. 
In general, the SD ferroelectric polarization can be written as  
\(
\boldsymbol{P}_\text{SD} = \sum_{i} \boldsymbol{P}_s(\boldsymbol{S}_i) + \sum_{\langle ij \rangle} \boldsymbol{P}_p(\boldsymbol{S}_i, \boldsymbol{S}_j),
\)  
where \(\boldsymbol{P}_s(\boldsymbol{S}_i)\) represents the spin–orbit coupling (SOC) mediated single-site term and \(\boldsymbol{P}_p(\boldsymbol{S}_i, \boldsymbol{S}_j)\) represents the pair term.
Based on our calculations, the SOC-induced contribution (\(\boldsymbol{P}_{\mathrm{SOC}} = 0.002~\mu\text{C/cm}^2\)) accounts for only about 5\% of the total spin-driven polarization. Therefore, the associated single-site term can be considered negligible, consistent with previous findings for BiFeO$_3$ and LaMn$_3$Cr$_4$O$_{12}$~\cite{xiang2013unified,feng2016anisotropic}.
We then concentrate on the pair terms \(\boldsymbol{P}_p\), which can be expressed as:
\begin{equation}
\begin{aligned}
\boldsymbol{P}_{p}(\boldsymbol{S}_{i}, \boldsymbol{S}_{j})& = \sum_{\alpha \beta} \boldsymbol{P}^{\alpha\beta}_{ij} \boldsymbol S_{i\alpha} \boldsymbol S_{j \beta} = \boldsymbol{S}_{i}^{T} \textbf{M}_{p} \boldsymbol{S}_{j}\\
&=  S_{ix} \boldsymbol{P}^{xx}_{ij} S_{jx} + S_{ix} \boldsymbol{P}^{xy}_{ij} S_{jy} + S_{ix}\boldsymbol{P}^{xz}_{ij} S_{jz}
\\
&+ S_{iy} \boldsymbol{P}^{yx}_{ij} S_{jx} + S_{iy} \boldsymbol{P}^{yy}_{ij} S_{jy} + S_{iy} \boldsymbol{P}^{yz}_{ij} S_{jz}
\\
&+ S_{iz} \boldsymbol{P}^{zx}_{ij} S_{jx} + S_{iz} \boldsymbol{P}^{zy}_{ij} S_{jy} + S_{iz} \boldsymbol{P}^{zz}_{ij} S_{jz}
\\
&=\begin{pmatrix} S_{ix},S_{iy},S_{iz} \end{pmatrix}
\begin{pmatrix} \boldsymbol{P}^{xx}_{ij} & \boldsymbol{P}^{xy}_{ij} & \boldsymbol{P}^{xz}_{ij} \\ \boldsymbol{P}^{yx}_{ij} & \boldsymbol{P}^{yy}_{ij} & \boldsymbol{P}^{yz}_{ij} \\ \boldsymbol{P}^{zx}_{ij} & \boldsymbol{P}^{zy}_{ij} & \boldsymbol{P}^{zz}_{ij} \end{pmatrix}
\begin{pmatrix} S_{jx} \\ S_{jy} \\ S_{jz} \end{pmatrix}
\label{spin-pair}
\end{aligned}
\end{equation}
 where \( \alpha, \beta \in \{x, y, z\} \). The ME tensor \( \mathbf{M}_p \) is a third-order tensor, with each element representing a vector \( \boldsymbol{P}_{ij}^{\alpha \beta} = ([P_{ij}^{\alpha \beta}]_x, [P_{ij}^{\alpha \beta}]_y, [P_{ij}^{\alpha \beta}]_z) \).

\begin{figure}[t]
\centering
	\includegraphics[scale=0.22]{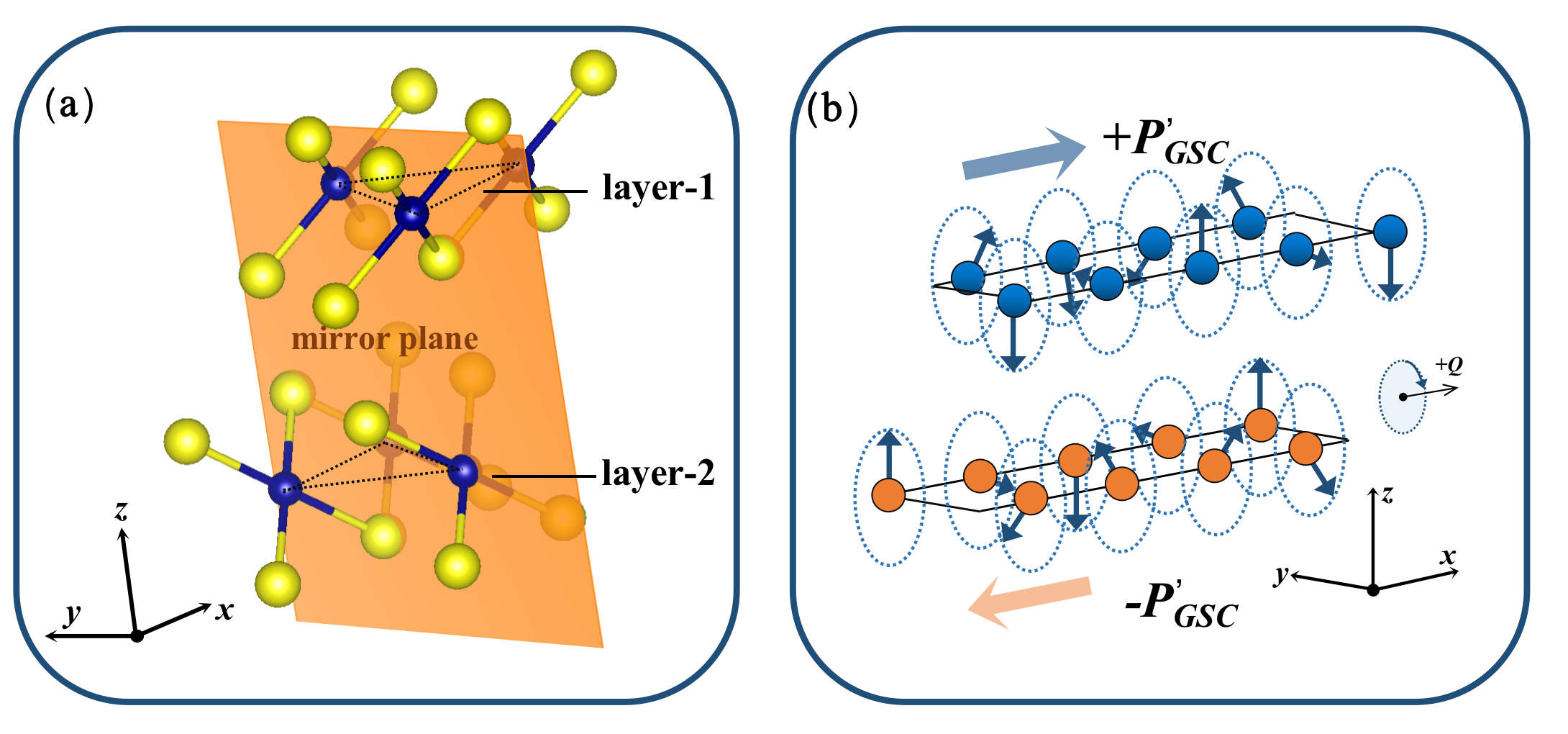}
	\caption{ \textbf{Representative double layers and local ferroelectric polarization in each layer.} (a)A mirror plane (\(yz\) plane) is perpendicular to the intrachain bond(along \(x\)-axis) and passes through its midpoint. The intralayer interchain bonds belong to the \(C_1\) point group.
(b) Local \(\boldsymbol{P}^{\prime}_{GSC}\) in each layer and their cancellation in two layers.}
\label{SYMMETRY ANALYZE}
\end{figure}

For simplicity, we only consider NN spin pairs for each bond type listed in Table~\ref{tab2}.
For the intraplane bonds, each spin site \(\mathbf{S}_{i}\) is coupled to six in-plane NN spins, corresponding to exchange paths of J\(_1\) and J\(_{1 \parallel}\) as shown in Fig.~\ref{Jpath}(a).
It is worth noting that the HM spin structure repeats every two layers. Therefore, for each \(\mathbf{S}_{i}\), two NN interlayer bonds between layer-1 and layer-2 are considered, which correspond to the exchange paths of J\(_{1 \perp}\), as shown in Fig.~\ref{Jpath}(b) and Fig.~\ref{SYMMETRY ANALYZE}(a).
As a result, the local SD polarization for each spin site \(\mathbf{S}_{i}\) can be expressed as:  
\(
\mathbf{P}_{i} = \frac{1}{2} \sum_{k=1}^{8} \mathbf{P}_{p}(\mathbf{S}_{i}, \mathbf{S}_{k}) = \frac{1}{2} \sum_{k=1}^{8} \mathbf{S}_{i}^{T} \mathbf{M}_{p}^{ik} \mathbf{S}_k
\),
where \( \mathbf{M}^{ik}_{p} \) represents the intersite ME tensor for the spin pair (\(i\), \( k \)).
The form of each ME tensor is determined by the local symmetry of each bond connecting the spin pairs (see Fig.~\ref{SYMMETRY ANALYZE}) [for a comprehensive derivation, see Ref.~\cite{xiang2013unified,zhang2017magnetic}]. These symmetry-imposed constraints on the ME tensors can be further validated by our numerical results calculated by the four-state mapping method. 
We then derive the uniform local SD polarization in the HM phase with \(\boldsymbol{Q} = (0.25, 0, 0)\), with their explicit expressions summarized in Table~\ref{formula}.
\begin{table}[t!]
    \centering
    \small
    \caption{ Local SD polarizations.  
Here, \(P^{\alpha \beta}\), \(P^{\alpha \beta}_{\parallel}\), and \(P^{\alpha \beta}_{\perp}\) denote tensor elements of the \(\boldsymbol{P}_{SD}\), categorized by bond types into intrachain, intraplane–interchain, and interlayer contributions, respectively.
}
    \renewcommand{\arraystretch}{1.6}
    \setlength{\tabcolsep}{5pt} 
	\begin{ruledtabular} 
    \begin{tabular}{cc}
    \toprule
$\mathbf{P}_{i}$ & Formula\\
    \hline
    $[\mathbf{P}_{i}]_{x}$ & \scalebox{0.8}{$\displaystyle [P^{yz}]_{x} +\frac{\sqrt{2} }{2} \big( [P_{\parallel}^{zy}]_{x}-[P_{\parallel}^{yz}]_{x} \big) -\frac{\sqrt{2}}{4}\big([P_{\perp}^{yz}]_{x}-[P_{\perp}^{zy}]_{x}\big)$} \\
    $[\mathbf{P}_{i}]_{y}$ & \scalebox{0.8}{$\displaystyle \frac{1}{2} \big( [P^{yz}]_{y} - [P^{zy}]_{y} \big) + \frac{\sqrt{2}}{4} \big( [P_{\perp}^{yy}]_{y} + [P_{\perp}^{zz}]_{y} - 2 [P_{\parallel}^{yy}]_{y} - 2 [P_{\parallel}^{zz}]_{y} \big)$} \\
    $[\mathbf{P}_{i}]_{z}$ & \scalebox{0.8}{$\displaystyle \frac{1}{2} \big( [P^{yz}]_{z} - [P^{zy}]_{z} \big) + \frac{\sqrt{2}}{4} \big( [P_{\perp}^{yy}]_{z} + [P_{\perp}^{zz}]_{z} - 2 [P_{\parallel}^{yy}]_{z} - 2 [P_{\parallel}^{zz}]_{z} \big)$} \\
 \toprule
    \end{tabular}
	\end{ruledtabular}
    \label{formula}
\end{table}
We here present the calculated values of the considered ME tensors, with all components given in units of \(10^{-5}\ \text{e}\cdot\text{Å}\). The corresponding bonds are classified by layers illustrated in Fig.~\ref{SYMMETRY ANALYZE}(a).
For the intrachain bond in layer-1:
\begin{equation}
    \begin{pmatrix}
(0,11,10)    &(5,93,-48)&(-4,-35,102)\\
  (5,-93,-48)  &(0,13,-2)&(75,0,3)\\
  (-4,35,-102)  &(-75,0,3)&(0,13,7).
    \end{pmatrix}
    \label{a-intrachian}
\end{equation}

For the intralayer interchain bond in layer-1:
\begin{equation}
\begin{pmatrix}
(0,\,-4,\,31) & (-1,\,-165,\,9) & (0,\,-127,\,40) \\
(-1,\,165,\,-9) & (0,\,-8,\,31) & (-31,\,-3,\,8) \\
(0,\,127,\,-40) & (31,\,-3,\,8) & (0,\,-7,\,31)
\end{pmatrix}
\label{me_a1}
\end{equation}

For the intrachain bond in layer-2:
\begin{equation}
\begin{pmatrix}
(0,\,-11,\,10) & (-5,\,93,\,48) & (-4,\,35,\,102) \\
(-5,\,-93,\,-48) & (0,\,-13,\,-2) & (-75,\,0,\,-3) \\
(-4,\,-35,\,-102) & (75,\,0,\,-3) & (0,\,-13,\,7)
\end{pmatrix}
\label{me_a2}
\end{equation}

For the intralayer interchain bond in layer-2:
\begin{equation}
\begin{pmatrix}
(0,\,4,\,31) & (1,\,-165,\,-9) & (0,\,127,\,40) \\
(1,\,165,\,9) & (0,\,8,\,31) & (31,\,-3,\,-8) \\
(0,\,-127,\,-40) & (-31,\,-3,\,-8) & (0,\,7,\,31)
\end{pmatrix}
\label{me_a3}
\end{equation}

For the interlayer bond:
\begin{equation}
\begin{pmatrix}
(0,\,0,\,-285) & (13,\,-6,\,-7) & (-6,\,2,\,41) \\
(-14,\,6,\,-6) & (0,\,0,\,-270) & (-4,\,0,\,-23) \\
(-6,\,2,\,-36) & (-4,\,0,\,24) & (0,\,0,\,-270)
\end{pmatrix}
\label{me_a4}
\end{equation}

By substituting these numerical values of ME tensors into the expressions in Table~\ref{formula}, we obtain a $\boldsymbol{P}_\text{SD}$ = (0, 0, -0.040) \(\mu\text{C/cm}^2\), which closely matches the result from our direct DFT calculations via the MSM approach.
In addition, our ME tensor analysis reveals that the NN interlayer bonds make the dominant contribution to $\boldsymbol{P}_\text{SD}$, accounting for approximately 78\% of the total.

~\\
\noindent {\bf\large Microscopic origin of the SD Polarization}\\

We now explore the microscopic origin of the SD polarization with direct DFT calculations. 
Magnetoferroelectricity typically arises from mechanisms such as exchange-striction (ES) (P $\propto$ \(\mathbf{S}_i \cdot \mathbf{S}_j\)) \cite{wang2016microscopic,lee2015giant}, which does not involve SOC and the SOC-facilitated generalized spin current (GSC) mechanism (P $\propto$ \(\mathbf{S}_i \times \mathbf{S}_j\)) \cite{xiang2011general} and higher-order \(p\)-\(d\) hybridization\cite{arima2007ferroelectricity}. 
Qualitatively, the ES polarization should be nonzero due to the lack of inversion symmetry in the \(Cmc2_1\) space group. Since the ES contribution does not depend on SOC, we perform direct DFT calculations without SOC to identify it. Our results suggest that the ES polarization
(\(\boldsymbol{P}_{ES} = (0, 0, -0.039)~\mu\text{C/cm}^2\)) dominates the overall SD polarization. As aforementioned, reversing the magnetic vector \(\mathbf{Q}\) within the same stacking pattern does not alter the total polarization.
Therefore, from a macroscopic perspective, this observation allows us to exclude contributions from the GSC mechanism and the higher-order \(p\)-\(d\) hybridization, which couple spin chirality to electric polarization.
Consequently, these SOC-mediated mechanisms, which typically account for SD polarizations in helical magnets such as NiI\(_2\)~\cite{yu2025microscopic,ni2025plane} and MnI\(_2\)~\cite{xiang2011general}, cannot account for the negligible SOC-induced polarization (\(P_{SOC} =(0, 0, -0.002)~\mu\text{C/cm}^2\)) in the $O$-phase CrI$_2$, which may instead originate from single-site mechanisms\cite{murakawa2010ferroelectricity,yamauchi2011theoretical}.

We next proceed with the analysis of the ME tensor to further investigate the microscopic mechanism.
In general, \( \mathbf{M}_p \) can be decomposed into three terms~\cite{wang2016microscopic,xiang2013unified,yu2025microscopic}: 
\(
\mathbf{M}_p = \mathbf{M}_{ES} + \mathbf{M}_{GSC} + \mathbf{M}_{AS},
\)
where the isotropic exchange-striction term is 
\(
\mathbf{M}_{ES} = \frac{1}{3} \text{Tr}(\mathbf{M}_p),
\)
the general spin-current term is 
\(
\mathbf{M}_{GSC} = \frac{1}{2}(\mathbf{M}_p - \mathbf{M}_p^T),
\)
and the anisotropic symmetric (AS) term is 
\(
\mathbf{M}_{AS} = \frac{1}{2}(\mathbf{M}_p + \mathbf{M}_p^T) - \mathbf{M}_{ES} \cdot \mathbf{I}.
\)
Here, \( \text{Tr} \) and the superscript \( T \) denote the trace and transpose of the \( \mathbf{M}_p \) tensor, respectively. \( \mathbf{I} \) is the identity matrix.
This decomposition further enables the systematic identification of local microscopic mechanisms of ME polarization.
By combining the expression for \(\mathbf{M}_{\mathrm{ES}}\) with the computed ME tensor [see Eqs.~(\ref{a-intrachian})-(\ref{me_a4})], our analysis indicates that the total SD polarization is indeed dominated by the ES effect,  in good agreement with our results from direct DFT calculations.

Importantly, the ME tensors reveal local GSC contributions \(\boldsymbol{P}^{\prime}_{\mathrm{GSC}}\) within each layer, arising from local symmetry breaking induced by the helical spin order, as illustrated in Fig.~\ref{SYMMETRY ANALYZE}(b). The $ \boldsymbol{P}^{\prime}_{\mathrm{GSC}}$ are calculated to be (\(0.0048, 0, 0 )\, \mu\text{C/cm}^2\) from the ME tensors. However, the \(\boldsymbol{P}^{\prime}_{\mathrm{GSC}}\) from the upper and lower layers cancel each other out, leading to a net-zero GSC contribution at the global scale.
To quantitatively evaluate the \(\boldsymbol{P}^{\prime}_{\mathrm{GSC}}\) from direct DFT calculations, we consider a virtual magnetic configuration, in which adjacent layers possess opposite magnetic chirality.
In this case, the direct DFT calculations yield a notable \(\boldsymbol{P}^{\prime}_{\mathrm{GSC}} \)=(0.0039, 0, 0)$\mu \text{C/cm}^2$, which is in good agreement with the value obtained from the ME tensor, thereby reinforcing the validity of our ME tensor analysis.

 ~\\
\noindent {\bf\large Monolayer CrI$_2$}\\

Given the predicted nonzero local \(\boldsymbol{P}^{\prime}_{\mathrm{GSC}}\) in each single layer of the bulk \(O\)-CrI\(_2\), we now turn our attention to monolayer CrI\(_2\) to verify whether this \(\boldsymbol{P}^{\prime}_{\mathrm{GSC}}\) persists in the monolayer limit.  In several previous theoretical studies, monolayer CrI\(_2\) was simply assumed to adopt the \(P\bar{3}m1\) space group~\cite{widyandaru2025tunable,yang2022interfacial}, which is commonly found in monolayers of transition metal dihalides\cite{botana2019electronic,kulish2017single}.
In this work, we assume that the CrI\(_2\) monolayer can be obtained by exfoliating a single layer from the bulk \(O\)-CrI\(_2\), which crystallizes in the \(Cm\) space group. The resulting CrI\(_2\) monolayer exhibits dynamical structural stability, as revealed by our calculated phonon spectrum [see Fig.2 in Supplementary Material].

To identify the magnetic ground state of monolayer CrI$_2$, we calculate the exchange parameters $J$'s. As summarized in Table~S2, the nearest-neighbor intrachain interaction $J_1$ switches from AFM to FM due to structural relaxation, where the Cr--I--Cr bond angle increases from $90.12^{\circ}$ (bulk) to $91.35^{\circ}$ (monolayer). Despite this, magnetic frustration along the $a$-axis persists due to competing $J_1$ and $J_2$, while other interactions remain largely unchanged.
The condition $\frac{|J_2|}{|J_1 + J_1^{\parallel}|} > \frac{1}{4}$ is still met, suggesting an HM ground state. Compared to the bulk, Monte Carlo simulations confirm that the ground state of the monolayer is still an HM, but with a slightly lower $T_N \sim 16$ K and a distinct propagation vector $\mathbf{Q}_\mathrm{m} = (0.125, 0, 0)$.
Total energy calculations including SOC show that the cycloid and proper-screw configurations are nearly degenerate, with an energy difference of $\sim 0.03$~meV/atom. For comparison, we also consider the bulk-like proper-screw HM order with $\mathbf{Q} = (0.25, 0, 0)$.

\begin{table}[t!]
    \centering
    \small
    \caption{Calculated \(\boldsymbol{P}^{\prime}_{\mathrm{GSC}}\) for the three magnetic configurations considered in the monolayer CrI$_2$.
}
    \renewcommand{\arraystretch}{1.6}
    \setlength{\tabcolsep}{5pt} 
	\begin{ruledtabular} 
    \begin{tabular}{cc}
    \toprule
Magnetic configuration &  \(\boldsymbol{P}^{\prime}_{\mathrm{GSC}}\) ($10^{-10}~\mu\text{C/cm}^2$ )\\
    \hline
    proper-screw with \(\mathbf{Q}_\text{m} = (0.125, 0, 0)\) & (9.45, 0.00, 0.00) \\  
     proper-screw with \(\mathbf{Q} = (0.25, 0, 0)\)& (1.22, 0.00, 0.00) \\
    cycloid with \(\mathbf{Q}_\text{m} = (0.125, 0, 0)\) &  (0.00, 9.78, 2.38)  \\
 \toprule
    \end{tabular}
	\end{ruledtabular}
    \label{PGSC}
\end{table}

The calculated \(\boldsymbol{P}^{\prime}_{\mathrm{GSC}}\) for the three magnetic configurations in monolayer CrI\(_2\) is summarized in Table~\ref{PGSC}. As expected, the proper-screw HM states produce polarization only along the \(x\)-axis, with the value for the predicted \(\boldsymbol{Q}_\mathrm{m}\) about one order of magnitude larger than that for the bulk-like \(\boldsymbol{Q}\).
In the \(Cm\) space group, the mirror plane perpendicular to the \(x\)-axis remains intact, constraining the intrinsic structural polarization \(\boldsymbol{P}_{\mathrm{in}}\) to the \(yz\)-plane. For reference, \(\boldsymbol{P}_{\mathrm{in}} = (0, -1723.94, 3.40) \times 10^{-10}~\mu\text{C/cm}^2\) is obtained from nonmagnetic calculations.
Since \(\boldsymbol{P}^{\prime}_{\mathrm{GSC}}\) reverses sign with magnetic chirality, electric fields along the \(x\)-axis are expected to enable chirality switching in the proper-screw state. In contrast, the cycloidal configuration yields a \(\boldsymbol{P}^{\prime}_{\mathrm{GSC}}\) in the \(yz\)-plane, which is masked by the dominant \(\boldsymbol{P}_{\mathrm{in}}\). Therefore, the cycloidal chirality may not be switchable by electric field in the $yz$ plane.

~\\
\noindent{\bf\large{Conclusions}}\\

We present a comprehensive first-principles study of magnetoelectric coupling in the vdW $O$-phase CrI$_2$. Monte carlo simulations based on DFT-derived exchange interactions confirm a helimagnetic ground state with a  N\'{e}el temperature consistent with experimental values. A ferroelectric switching pathway facilitated by interlayer sliding is proposed, featuring a low switching energy barrier and out-of-plane ferroelectric polarization.
To quantify the spin-driven polarization \(\boldsymbol{P}_{\mathrm{SD}}\), we employ two distinct approaches: direct comparison between the helimagnetic and paramagnetic phases, and evaluation using the magnetoelectric tensor. Both approaches yield consistent results, showing \(\boldsymbol{P}_{\mathrm{SD}}\) along the \(z\)-axis, primarily originating from the exchange-striction mechanism.
Although the net \(x\)-polarization vanishes in the bulk due to symmetry constraints, each CrI\(_2\) layer exhibits a local in-plane polarization via the GSC mechanism, which directly couples to the magnetic chirality. 
We further predict that, in monolayer CrI\(_2\), the magnetic chirality can be switched by reversing the in-plane polarization under an external electric field, offering a promising route toward electrical manipulation of magnetic chirality in two-dimensional multiferroics.

~\\
\noindent{\bf\large Methods}

\noindent{\bf DFT + $U$ calculations.} 
Our first-principles calculations employ the Vienna $ab\ initio$ Simulation Package (VASP)~\cite{vasp-1,vasp-2}. The exchange-correlation functional is treated using the spin-polarized Perdew-Burke-Ernzerhof generalized gradient approximation (GGA). We utilize a $\Gamma$-centered 16 × 8 × 4 k-point mesh. The projector augmented wave (PAW) method~\cite{PAW} with a 500 eV plane-wave cutoff energy is employed. To properly describe the Cr 3$d$ orbitals, we apply the GGA+$U$ formalism~\cite{GGA+U} with Hubbard parameters $U$ = 3 eV and $J$ = 1 eV (yielding effective $U_{eff}$  = 2.0 eV), which well reproduces experimental magnetic moments and magnetic transition temperatures~\cite{PhysRevB.109.144403}. Experimental lattice parameters are used throughout our calculations, with atomic positions relaxed until residual forces below 1 meV/\AA. 
\\

\noindent{\bf Heisenberg exchange interactions and MC calculations.}
Magnetic exchange parameters $J$'s are determined through fitting to energies from randomly generated collinear magnetic configurations~\cite{ni2025plane,PhysRevB.105.214303}. Magnetic phase diagrams are obtained using replica-exchange MC simulations~\cite{cao2009first}. 
The MC simulations are performed with a 12 $\times$ 12 $\times$ 3 superlattice with 1 000 000 MC steps for statistics at each temperature.
\\

\noindent{\bf Electric polarization and ME tensor calculations} 
Electric polarization is calculated via the Berry phase formalism~\cite{berryphase}. All ME tensor $\text{M}_{p}$ is evaluated via the four-state mapping method (no-substitution)~\cite{xiang2011general}.
Specifically, the inherent non-spin-driven polarization remains constant across the four states and is effectively eliminated by pairwise combination and subtraction, thereby isolating the SD polarization.
The ferroelectric switching pathway is obtained by using the climbing image nudged elastic band (CI-NEB) method\cite{henkelman2000climbing}.
\\

~\\
\noindent {\bf\large Data availability}\\
All data needed to evaluate the conclusions in the paper are available within the article. All raw data generated during the current study are available from the corresponding author upon request.

~\\
\noindent {\bf\large Code availability}\\
The codes used for the DFT calculations in this study are
available from the corresponding authors upon  request.

~\\
\noindent {\bf\large References}

\vspace{3mm}

\noindent {\bf\large Acknowledgements}\\
This work was supported by the National Key R$\&$D Program of China (Grant No. 2023YFB4603801), National Natural Science Foundation of China (Grant No.12474249),
Guangdong Provincial Key Laboratory of Magnetoelectric Physics and Devices (No. 2022B1212010008).
This work is supported in part by The Major Key Project of Peng Cheng
Laboratory (PCL).	
\vspace{3mm}

\noindent {\bf\large Author Contributions}\\
H.S.Y and K.C. proposed and designed the research. H.S.Y and X.S.N. contributed to the DFT, MC and electric polarization calculations.  H.S.Y, X.S.N, and K.C analyzed the data. H.S.Y, X.S.N. and K.C. wrote the paper. All authors participated in the discussion and comment on the paper.

\vspace{3mm}

\noindent{\bf\large Competing interests}\\
The authors declare no competing interests.

\appendix 
\setcounter{table}{0} 
\setcounter{figure}{0}
\setcounter{section}{0}
\setcounter{equation}{0}
\renewcommand{\thetable}{A\arabic{table}}
\renewcommand{\thefigure}{A\arabic{figure}}
\renewcommand{\thesection}{A\arabic{section}}
\renewcommand{\theequation}{A\arabic{equation}}

\end{document}